\documentclass[USenglish]{article}
\usepackage[utf8]{inputenc}
\usepackage[big,online]{dgruyter}
\setcitestyle{numbers,sort&compress}

\newcommand\nc{\newcommand*}  \nc\longnc{\newcommand}

\longnc\OMIT[1]{\relax}  
\longnc\VOMIT[1]{#1}     

\nc\eq[2]{\begin{align}\label{#1}#2\end{align}}

\nc\re[1]{(\ref{#1})}
\nc\m[1]{$#1$}
\nc\qfrac[2]{{\textstyle \frac{#1}{#2}}}
\nc\qqfrac[2]{\9 1 {#1}\big/{#2}}

\nc\0[2]{\ifcase#1{#2}\or\lt(#2\rt)\or\lt[{#2}\rt]\or\lt\{{#2}\rt\}\or
  \mathord<{#2}\mathord>\or\lt\langle{#2}\rt\rangle\or\lt\lvert{#2}\rt
  \rvert\or\lt\lVert{#2}\rt\rVert\fi}
\nc\1[2]{\ifcase#1{#2}\or(#2)\or[#2]\or\{#2\}\or\mathord<{#2}\mathord
  >\or\langle{#2}\rangle\or\lvert{#2}\rvert\or\lVert{#2}\rVert\fi}
\nc\2[2]{\ifcase#1{#2}\or\big(#2\big)\or\big[#2\big]\or\big
  \{#2\big\}\or\big<#2\big>\or\big\langle#2\big\rangle\or\big
  \lvert#2\big\rvert\or\big\lVert#2\big\rVert\fi}
\nc\3[2]{\ifcase#1{#2}\or\Big(#2\Big)\or\Big[#2\Big]\or\Big\{#2\Big
  \}\or\Big<#2\Big>\or\Big\langle#2\Big\rangle\or\Big\lvert#2\Big
  \rvert\or\Big\lVert#2\Big\rVert\fi}
\nc\4[2]{\ifcase#1{#2}\or\bigg(#2\bigg)\or\bigg[#2\bigg]\or\bigg
  \{#2\bigg\}\or\bigg<#2\bigg>\or\bigg\langle#2\bigg\rangle\or\bigg
  \lvert#2\bigg\rvert\or\bigg\lVert#2\bigg\rVert\fi}
\nc\5[2]{\ifcase#1{#2}\or\Bigg(#2\Bigg)\or\Bigg[#2\Bigg]\or\Bigg
  \{#2\Bigg\}\or\Bigg<#2\Bigg>\or\Bigg\langle#2\Bigg\rangle\or\Bigg  
  \lvert#2\Bigg\rvert\or\Bigg\lVert#2\Bigg\rVert\fi}
\nc\9[2]{\ifcase#1{#2}\or\left(#2\right)\or\left[#2\right]\or\left
  \{#2\right\}\or\left\langle{#2}\right\rangle\or\left\langle{#2}\right
\rangle\or\left\lvert{#2}\right\rvert\or\left\lVert{#2}\right\rVert\fi}
\nc\lt{\mathopen{}\mathclose\bgroup\left}
\nc\rt{\aftergroup\egroup\right}

\nc\bi\relax                      
\nc\bit{      \mskip1mu}  \nc\biT{      \mskip-1mu}   
\nc\bitt{     \mskip2mu}  \nc\biTT{     \mskip-2mu}   
\nc\bittt{    \mskip3mu}  \nc\biTTT{    \mskip-3mu}   
\nc\bitttt{   \mskip4mu}  \nc\biTTTT{   \mskip-4mu}   
\nc\bittttt{  \mskip5mu}  \nc\biTTTTT{  \mskip-5mu}   
\nc\bitttttt{ \mskip6mu}  \nc\biTTTTTT{ \mskip-6mu}
\nc\bittttttt{\mskip7mu}  \nc\biTTTTTTT{\mskip-7mu}

\nc\tensUpRt[1]{^{\mathrm{#1}}}            
\nc\symm{\tensUpRt{S}}   \nc\asymm{\tensUpRt{A}}
\nc\dev{\tensUpRt{d}}    \nc\sph{\tensUpRt{s}}
\nc\QI{\rule{0em}{1.6ex}}
\nc\qinv[1]{\0 1 {{#1}^{-1}}\QI}
\nc\qsymm[1]{\0 1 {{#1}\symm}\QI}
\nc\tensor{}
\nc\qtensor\mathbf  
\nc\Tensor{}
\nc\four{}
\nc\qf{\hat}
\nc\Q{\mskip7.5 mu}
\nc\QQ{\mskip11.5 mu}

\nc\dd{\mathrm{d}}
\nc\pd\partial
\nc\qdot{^{\biTT\hbox{\boldmath{$\cdot$}}}}
\nc\qpar{{\scriptscriptstyle \|}}
\nc\qel[1]{{#1}_{\text{el}}}
\nc\qth[1]{{#1}_{\text{th}}}
\nc\qrh[1]{{#1}{}^{}_{\text{rheol}}}

\nc\qalp{\alpha}       \nc\qqalp{\Tensor{\qalp}}    \nc\qqqalp{\qE_0}
                                                    \nc\qqqbet{\qE_1}
\nc\qgam{\gamma}                                    \nc\qqqgam{\qE_2}
\nc\qdel{\delta}       \nc\qqdel{\Tensor{\qdel}}
\nc\qeps{\varepsilon}  \nc\qqeps{\Tensor{\qeps}}
\nc\qlam{\lambda}
\nc\qpi{\pi}
\nc\qrho{\varrho}
\nc\qsig{\sigma}       \nc\qqsig{\Tensor{\qsig}}
                       \nc\qqtau{\tau}
\nc\qxi{\xi}           \nc\qqxi{\Tensor{\qxi}}

\nc\qGam{\Gamma}

\nc\qa{a}              \nc\qqa{\tensor{\qa}}
\nc\qc{c}
\nc\qe{e}
\nc\qg{g}              \nc\qqg{\tensor{\qg}}
\nc\qh{h}              \nc\qqh{\tensor{\qh}}
\nc\qj{j}              \nc\qqj{\tensor{\qj}}
                                                    \nc\qqql{l}
\nc\qp{p}
\nc\qr{r}
\nc\qs{s}
\nc\qu{u}
\nc\qv{v}              \nc\qqv{\four{\qv}}
\nc\qw{w}

\nc\qA{A}              \nc\qqA{\tensor{\qA}}
\nc\qC{C}              \nc\qqC{\tensor{\qC}}
\nc\qD{D}              \nc\qqD{\tensor{\qD}}
\nc\qE{E}
\nc\qF{F}              \nc\qqF{\tensor{\qF}}
\nc\qH{H}
                       \nc\qqI{\tensor{I}}
\nc\qJ{J}              \nc\qqJ{\tensor{\qJ}}
\nc\qL{L}              \nc\qqL{\tensor{\qL}}
                       \nc\qqN{\tensor{0}}
\nc\qT{T}
\nc\qZ{Z}              \nc\qqZ{\tensor{\qZ}}

\nc\Qa{i}  \nc\Qb{j}  \nc\Qc{k}  \nc\Qd{l}  \nc\Qe{m}  \nc\Qf{n}
\nc\QA{I}  \nc\QB{J}  \nc\QC{K}  \nc\QD{L}  \nc\QE{M}  \nc\QF{N}

\begin{document}

\OMIT{
  \articletype{Research Article}
  \journalname{J.~Non-Equilib.~Thermodyn.}
  \journalyear{2015}
  \journalvolume{???}
  \journalissue{???}
  \startpage{1}
  \aop
  \DOI{10.1515/jnet-YYYY-XXXX}
}

\title{Objective thermomechanics}

\author[1]{Tam\'as F\"ul\"op}

\affil[1]{\protect\raggedright 
 Department of Energy Engineering, Budapest University of Technology and
 Economics, 1111 Budapest, M\H uegyetem rkp. 3.; Montavid Thermodynamic
 Research Group, 1112 Budapest, Igm\'andi u. 26.; e-mail:
 fulop@energia.bme.hu}

\abstract{
An irreversible thermodynamical theory of solids is presented  where the
kinematic quantities are defined in an automatically objective way.
Namely, auxiliary elements like reference frame, reference time and
reference configuration are avoided by formulating the motion of the
continuum on spacetime directly. Solids are distinguished from fluids by
possessing not only an instantaneous metric tensor but also a relaxed
metric. The elastic state variable is defined through comparing these
two metrics. Thermal expansion is conceived as temperature
dependence of the relaxed metric and plasticity, an irreversible
change in the relaxed metric, is described via a plastic change rate
tensor. Thermomechanics is built around these -- finite deformation --
kinematic quantities by starting from mechanics and adding
thermodynamical requirements gradually. The obtained theory is not
restricted to isotropic media.
 }





\maketitle

\section{Introduction}\label{intro}

In Euclid's geometry, one operates with points, lines, triangles,
vectors, rotations, reflections, and there may seem no need for
coordinates. Descartes' suggestion to use coordinates comes convenient
for applicational calculations for complicated geometric objects. It is
inconvenient, on the other side, for principles and understanding.
Nevertheless, along the many successful applications, coordinates became
the standard language of geometric description in physics and related
areas.

Reference frames are analogous objects on spacetime. The Galilean
principle of relativity -- the equivalence of \1 1 {inertial} reference
frames -- unfolds not only that motion is not absolute but also that
space is not absolute. Namely, in terms of the Galilean transformation
rule in customary notation, while time is absolute, \m { t' = t = f(t)},
space is not: \m { \qtensor{r}' = \qtensor{r} - \qtensor{V}t \ne
\qtensor{g}(\qtensor{r}) }: space is inevitably intertwined with time.

When Newton created his dynamics, he assumed an absolute space
because he knew no other way to formulate his action-at-a-distance type
description of gravitation -- neither mathematics nor physics was
developed enough to provide him a more appropriate framework. Along the
success of his dynamics, absolute space also became a standard part of
mechanics.

Later, time proved relative as well, and the Galilean transformation
rule has been superseded by the Lorentz transformation rule:
 \eq{@14071}{
t' = \qqfrac {t - \qfrac {\qtensor{V}}{c^2}\qtensor{r}_{\qpar}} {\sqrt{1
- V^2/c^2}} ,
 \qquad
\qtensor{r}'_{\qpar} = \qqfrac {\qtensor{r}_{\qpar} - \qtensor{V}t}
{\sqrt{1 - V^2/c^2}} ,
 \qquad
\qtensor{r}'_{\perp} = \qtensor{r}^{}_{\perp} .
 }
When, correspondingly, Einstein formulated his relativity theory, he
spoke in terms of reference frames, and about their relationships. Along
the success of his ideas, reference frames also became the de facto
standard when treating spacetime.

Nevertheless, Weyl \cite{Wey18}, and later independently Matolcsi
\cite{Mat84b,Mat93b}, have pointed out that  a reference frame free
description of spacetime is possible. According to this, in both the
Galilean and the special relativistic case, spacetime is a four
dimensional affine space, equipped by some further structure: In the
Galilean case%
, an absolute time structure (foliation,
``slicing'') and a Euclidean structure on the equal-time subspaces
(``slices''), while in the special relativistic case a Lorentz metric
(pseudo-Euclidean form).

\begin{figure}[ht]
 \begin{center}
 \hfill
\includegraphics[width=.33\columnwidth]{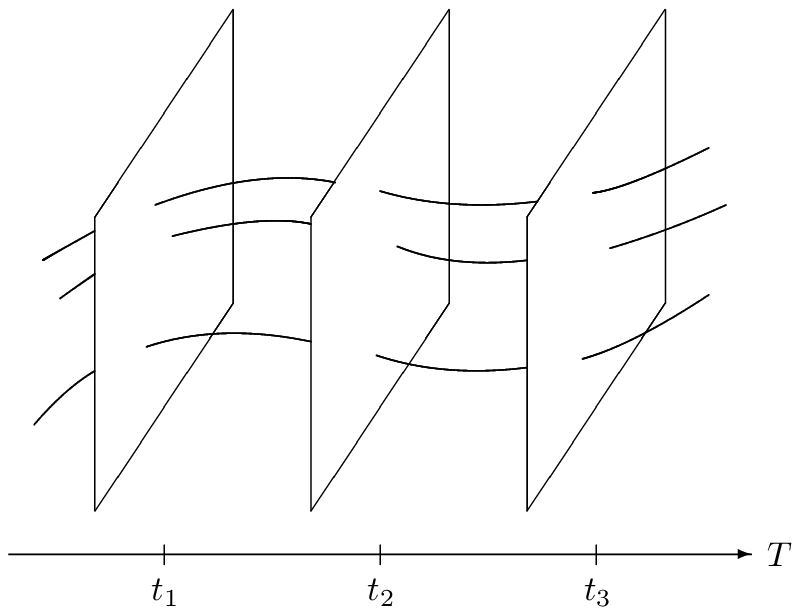}
 \hfill
\includegraphics[width=.33\columnwidth]{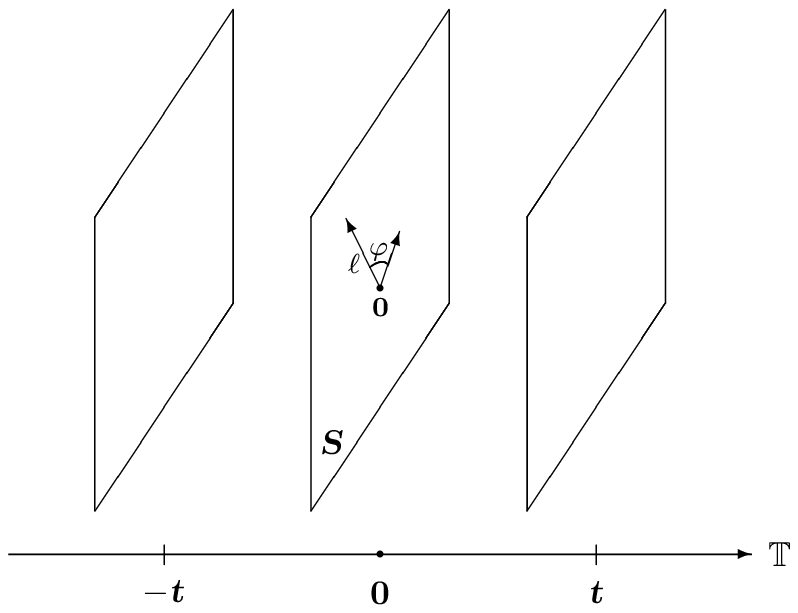}
 \hfill\null
\caption{Left: three world lines in Galilean spacetime; right: Galilean
spacetime vectors}
 \label{fbf}
\end{center}
\end{figure}

In the continuum thermodynamics (thermomechanics, etc.) of solids,
various customary auxiliary elements are used, like a reference frame --
including a coordinate system --, an initial/reference time \m { t_0 },
a reference configuration (the distribution of material points at
reference time in the space of the reference frame), and a constant
temperature \m { \qT_0 } at reference time. Objectivity is the
requirement that there must be a physical content behind our
description, which content is independent of our description. The
customary formulation of objectivity is telling that, when changing
auxiliary elements, what transforms how.

This approach has turned out to lead to controversies, artifacts and
mistakes. Moreover, the essence (of the phenomena) is hidden behind the
formulation (via the auxiliary elements). The situation is analogous to
electromagnetism where, for more easily solvable equations, it is
customary to replace the field strength four-tensor by a four-potential.
The four-potential is not unique, there is a freedom -- the so-called
gauge freedom -- in its choice so a gauge fixing condition is required.
This condition does not belong to the physical phenomenon but to this
type of its description. While \textit{formally} convenient,
\textit{physically} the four-potential is misleading, which is revealed,
for example, in approximations like perturbative solutions, where the
smallness of a perturbative correction turns out to depend heavily on
the gauge fixing condition.

The term `auxiliary' is, in the present context, related to the meanings
`something  nonessential, arbitrary, and misleading'. A direct, i.e.,
auxiliary element free, spacetime description of continuum
thermodynamics of solids would have various advantages. First,
objectivity would automatically be guaranteed. Next, essence would be
more directly visible (what causes what, etc.). Quantities which one
ordinarily considers as scalars, three-vectors or three-tensors could be
realized as relative components of absolute spacetime four-quantities:
four-scalars, four-vectors or four-tensors. Auxiliary elements would be
applied only in concrete applications -- calculating a given process of
a given sample of given properties --, they should be introduced only to
the necessary amount, and should be used with one eye continuously on
what distortions those auxiliary elements may cause on our physical
picture of the phenomenon.

However, if one uses no reference frame, no reference time and no
reference configuration then one has no displacement field \m {
\qtensor{u} }, no deformation gradient \m { \qtensor{F} }, no strain
tensor \m { \qqeps } nor any customary deformation measure: Is it then
possible to formulate elasticity? Is it possible to formulate
plasticity? If one has no initial temperature \m { T_0 } then how to
formulate thermoelasticity?

The task of this paper is to show that, yes, everything necessary can be
formulated. Partial answers have already been published in recent works:
The spacetime-friendly elastic and plastic kinematic quantities for
solid continua have been introduced in \cite{FulVan10}, thermal
expansion was added in \cite{Ful11}, and thermomechanics in the
small-strain regime and for isotropic materials was presented and used
for evaluating experimental data in
\cite{FulVanCsa13Bre,FulVanCsa13Bal,AssCsaFul15}. The present paper
extends the treatment to large-deformation thermomechanics and to
anisotropic materials.

The spacetime perspective is quite an abstract one, while the impression
`this theory is so abstract that it cannot be used for concrete
situations' should be avoided. Therefore, thermomechanics is built here
step by step starting from mechanics and in terms of experimentally
readily accessible quantities. This attitude also helps in ensuring
thermodynamical consistency. Namely, one could postulate thermodynamical
potentials and the Gibbs relation but, historically, the Gibbs relation
was born in the context of gases, and who knows a priori how to
generalize it for large deformations of anisotropic solid bodies, with
their tensorial -- and therefore not necessarily commuting --
quantities? During the present step-by-step approach, one can see what
one can have and how one can ensure positive entropy production in the
end. 

This paper treats the case of Galilean spacetime model only. {Formally,}
there seems no obstacle to generalize the obtained formulae for special
and general relativistic background. {Physically}, however, many
constitutive assumptions may break. For example, elasticity has been
born to express a type of deviation from rigid body behaviour, and
assumes internal forces depending on the equal-time distances among
material points, but special/general relativistically there are no rigid
\1 1 {accelerated} bodies and `equal time' is a frame dependent notion.
Other interactions among the different parts of the continuum also
propagate with a finite speed so thermodynamical aspects are also
nontrivial to realistically generalize.

In the Galilean spacetime model, many notions -- spacetime points, world
lines, four-velocity etc.\ -- are analogous to those of special
relativistic spacetime theory so a knowledge of the latter helps a lot to
understand the former. Concerning reading background for the differences,
Weyl's book \cite{Wey18} does not work out the technical details for
practical applicability, and Matolcsi's English books
\cite{Mat84b,Mat93b} are not easy to find, but the Appendix of
\cite{FulVan10} provides a helpful introduction and summary.

\section{Kinematic quantities for solid continua}\label{kinem}

As the first step, a continuum is modelled by a three dimensional smooth
manifold, called hereafter the material manifold. \1 1 {Boundaries are
not treated in the present paper.} The tangent vectors of this manifolds
are referred to as material vectors. These give rise to, along usual
manifold theory, tensors \1 1 {called material tensors} of various type.
It is worth emphasizing that covectors \1 1 {real valued linear forms on a
tangent space} are not identified with vectors \1 1 {of that tangent
space}.

The existence of any material point \1 1 {i.e., point of the material
manifold} in spacetime is described by a world
line. This provides a system of world lines \m { \qr }: a given
material point \m { P } at a given time \m { t } is at spacetime point
\m { \qr \1 1 { t, P }}. Differentiation with respect to the time
variable will be denoted by overdot and in the material variable by
 \m { \nabla_{\QC} }.
Here and hereafter, Penrose's abstract index notation is used, where
indices do not refer to components with respect to a coordinate system
but to vectorial/tensorial type \1 1 {without the need for a coordinate
system}. Upper indices indicate vectors and lower ones covectors, and an
index appearing twice, once in upper and once in lower position,
indicates tensorial contraction. We need to treat various manifolds, and
the conventions applied are best explained on the example of the
world line gradient,
 \m{ \qqJ^{\Qc }_{\Q \QC} :=  \nabla_{\QC } \qr^{\qf \Qc} }
which is, at time \m { t }, the differential map from the material
manifold to the three dimensional flat Riemannian manifold of equal-\m {
t } spacetime points, the metric of which is the Euclidean inner product
 \m { \qqh_{\Qa \Qb} }
of spacelike spacetime vectors. Namely, capital letters like \m { \QC }
here refer to material indices, small ones with overhat like \m { \qf
\Qc } to four-indices, i.e., indices for spacetime as a manifold, and
small ones without overhat like \m { \Qc } here to spacelike spacetime
vectors. Though no coordinate systems are assumed here, formulae
\textit{look like} ones with coordinate systems introduced, and most
calculational rules for coordinate indices are valid here, too. When it
is important to indicate that a certain formula does not hold in the
coordinate sense but only in the abstract sense, this will be denoted by
braces like in \m { \9 3 {\qqJ^{\Qc }_{\Q \QC}} }.

The derivative
 \m { \qqv^{\qf \Qc} := \dot\qr^{\qf \Qc} }
is the four-velocity field, the gradient of which defines the velocity
gradient
 \m { \qqL^{\Qc }_{\Q \QD} := \nabla_{\QD } \qqv^{\qf \Qc} =
 {\dot\qqJ}^{\Qc }_{\Q \QD} },
\,
 \m { \qqL^{\Qc }_{\Q \Qd} := \nabla_{\Qd } \qqv^{\qf \Qc} =
 {\dot\qqJ}^{\Qc }_{\Q \QC} \qinv{\qqJ}^{\QC }_{\QQ \Qd} }.
One can observe here an example of the general differential geometrical
role of the differential map \m { \qqJ^{\Qc }_{\Q \QC} }: It transports
material vectors to spacelike spacetime ones and vice versa, material
covectors to spacelike spacetime covectors, etc. -- depending on
context, its inverse or transpose or both may need to be used.

Another important notion is the instantaneous metric
 \m{ \qqh^{}_{\QC \QD} := \qqJ^{\Qc }_{\Q \QC} \qqh^{}_{\Qc \Qd}
 \qqJ^{\Qd }_{\Q \QD} },
which describes the instantaneous distance of any two material points.
This metric makes the material manifold a flat Riemannian manifold for
any instant. The instantaneous metric
is heavily process dependent, motion dependent, and one finds for its
time derivative
 \eq{@21811}{
\dot\qqh^{}_{\QC \QD} = \dot\qqJ^{\Qc}_{\Q \QC} \qqh^{}_{\Qc \Qd}
\qqJ^{\Qd }_{\Q \QD} + \qqJ^{\Qc}_{\Q \QC} \qqh^{}_{\Qc \Qd}
\dot\qqJ^{\Qd }_{\Q \QD} = \qqL^{\Qc}_{\Q \QC} \qinv\qqJ^{\QE}_{\Q\Q
\Qc} \qqh^{}_{\QE \QD} + \qqh^{}_{\QC \QE} \qinv\qqJ^{\QE}_{\Q\Q \Qd}
\qqL^{\Qd }_{\Q \QD} = \qqL^{\QE}_{\Q\Q
\QC} \qqh^{}_{\QE \QD} + \qqh^{}_{\QC \QE} \qqL^{\QE}_{\Q\Q \QD} .
 }

In the spirit of the general methodology of Matolcsi
\cite{Mat84b},
we intend to give a mathematical model to every physical notion (so
typically any notion gets modelled by either a set or a map from a set
to another). Our description so far has not distinguished
fluids from solids. How to formulate the difference?

Our everyday experience says that solids are ``solid''. As a zeroth
approximation, solids could be modelled by a rigid body. Naturally, for
elasticity we need to go beyond that level: Distances of an elastic body
are not time independent. Nevertheless, for any pair of material points,
there seems to be a distinguished distance, encoded somehow in the body,
which is the distance when the body is in an undisturbed and completely
relaxed state. This intuitive picture can be modelled in the following
way: In addition to the instantaneous metric, solids are equipped with a
relaxed metric
 \m{ \qqg^{}_{\QC \QD} },
too. In a relaxed state,
 \m{ \qqh^{}_{\QC \QD} = \qqg^{}_{\QC \QD} }.

In general,
 \m { \qqh^{}_{\QC \QD} \ne \qqg^{}_{\QC \QD} },
and the extent to which they differ is measured conveniently by the
elastic shape tensor
 \m{ \qqA^{\QC }_{\QQ \QD} := \qinv{\qqg}^{\QC \QE} \qqh_{\QE \QD} },
which is the unit tensor
 \m { \qqI^{\QC }_{\QQ \QD} }
in relaxed state and some other, but still nondegenerate, tensor
otherwise (its determinant is never zero). Historically, we are
accustomed to a deformation measure that is zero in relaxed state, and
to which Hookean elastic stress is proportional, i.e., which
generalizes Cauchy's small-strain tensor -- this explains why
the
Biot, Hencky, Almansi etc. tensors have been
defined.
Hence, the elastic deformedness tensor is also
practical to introduce:
 \m{ \9 3 { \qqD^{\QC }_{\QQ \QD} } := \frac{1}{2} \ln \9 3 { \qqA^{\QC
 }_{\QQ \QD} } . }
This logarithmic definition is a distinguished choice. Indeed, on one
side, it can be shown that the spherical part of this tensor,
 \m { \0 1 {\qqD\sph}\QI^{\QC }_{\QQ \QD} },
is zero for volume-preserving motions while its
deviatoric part,
 \m { \0 1 {\qqD\dev}\QI^{\QC }_{\QQ \QD} },
is zero for isotropic volumetric changes. Furthermore, it also gives
rise to the property
 \eq{@13637}{
\frac { \dd f }{\dd \qqD^{\QC }_{\QQ \QD}} =
2 \frac { \dd f }{\dd \qqA^{\QE }_{\QQ \QD}} \qqA^{\QE }_{\QQ \QC} =
2 \qqA^{\QD }_{\QQ \QF} \frac { \dd f }{\dd \qqA^{\QC }_{\QQ \QF}}
 }
for any isotropic scalar function \m { f }. Property \re{@13637} can be
proved by writing \m { f } as a function of the isotropic invariants of
 \m { \qqA^{\QC }_{\QQ \QD} },
and differentiating it as a composite function. For the isotropic
invariants themselves property \re{@13637} is easy to see. As a part of
this, why
 \m { \qqA^{\QC }_{\QQ \QD} }
commutes with the derivative of \m { f } with respect to it follows from
that the derivative of these isotropic invariants with respect to
 \m { \qqA^{\QC }_{\QQ \QD} }
are
 some powers of
 \m { \qqA^{\QC }_{\QQ \QD} },
multiplied by some scalar so commuting holds term by term.

As long as we remain in the range of elastic phenomena, the relaxed
metric is considered constant. Accordingly, one can
derive the following evolution equation for the spacetime version
 \m{ \qqA^{\Qa }_{\Q \Qb} = \qqJ^{\Qa }_{\Q \QC} \qqA^{\QC }_{\QQ \QD}
 \qinv{\qqJ}^{\QD }_{\QQ \Qb} } from \re{@21811}:
 \m{
{\dot{\qqA}}^{\Qa }_{\Q \Qb} = \qqL^{\Qa }_{\Q \Qc} \qqA^{\Qc }_{\Q \Qb}
+ \qqA^{\Qa }_{\Q \Qc} \qinv{\qqh}^{\Qc \Qd} \qqL^{\Q \Qe}_{\Qd }
\qqh^{}_{\Qe \Qb} .
 }

Also while staying within the range of elasticity, there seems no need to
expect that
 \m { \qqg^{}_{\QC \QD} }
is not flat. Correspondingly to flatness, a condition can be derived for
the Ricci tensor of the relaxed metric, which is the kinematic
compatibility condition for
 \m { \qqg^{}_{\QC \QD} }.
In the present, large deformation, description, the compatibility
condition proves to be a rather complicated formula for
 \m { \qqA^{\Qa }_{\Q \Qb} }
\cite{FulVan10}.
Its small-deformedness leading order is the well-known simple form of the
compatibility condition (left + right curl of
 \m { \0 0 { \qqD^{\Qa }_{\Q \Qb} } }
is zero).

As a consequence of systematically distinguishing covectors from
vectors, we can see that elastic shape and elastic deformedness are not
simply symmetric tensors. The proper -- and apparent -- property is
that they are \m { \qqh }-symmetric and \m { \qqg }-symmetric:
Combinations like
 \eq{@27287}{
 &&
\qqh^{}_{\QA \QB} \qqA^{\QB }_{\Q \QC} & = \qqh^{}_{\QA \QB}
\qinv{\qqg}^{\QB \QD} \qqh^{}_{\QD \QC} ,
 &
\qqA^{\QA }_{\Q \QB} \qinv{\qqg}^{\QB \QC} & = \qinv{\qqg}^{\QA \QD}
\qqh^{}_{\QD \QE} \qinv{\qqg}^{\QE \QC}
 &&
 }
  and the corresponding spacetime tensorial versions,
are symmetric, following from the symmetricity of \m { \qqh^{}_{\QA \QB}
}, \m { \qqh^{}_{\Qa \Qb} }, \m { \qqg^{}_{\QA \QB} } and \m {
\qqg^{}_{\Qa \Qb} }. Mixed tensors like \m { \qqA^{\QA }_{\Q \QB} } can
never be symmetric or antisymmetric. On the other side, only mixed
tensors can have a determinant, a trace and eigenvalues--eigenvectors,
such as \m { \qqA^{\QA }_{\bitttt \QB} \qqa^{\QB } = \qlam \qqa^{\QA } }.

Thermal expansion is a phenomenon which enforces us to go beyond
elasticity. In the language of the relaxed metric, thermal expansion can
be formulated by allowing
 \m{ \qqg^{}_{\QC \QD} = \qqg^{}_{\QC \QD} (\qT)}.
Accordingly,
 \eq{@17071}{
\qqalp^{\QQ \QB}_{\QA } := \frac {1}{2} \3 0 {\frac {\dd \qqg^{}_{\QA
\QC}}{\dd \qT} } \qinv{\qqg}^{\QC \QB} ,
 }
the thermal expansion coefficient tensor, can be defined. For isotropic
materials,
 \m { \qqalp^{\QQ \QD}_{\QC } = \qalp \qqI^{\QQ \QD}_{\QC } }.
On the other side, in anisotropic cases, the order of
the matrix product \re{@17071} is relevant -- why this product order is
preferred to
 \m { \frac {1}{2} \qinv{\qqg}^{\QA \QC} \2 0 {\frac {\dd \qqg_{\QC \QB}}{\dd
 \qT} } }
will become clear in Sect.~\ref{ani}.
 The relaxed metric now becoming process dependent, the following
generalization of the above kinematic time evolution equation can be found:
 \eq{@18040}{
{\dot{\qqA}}^{\Qa }_{\Q \Qb} = \qqL^{\Qa }_{\Q \Qc} \qqA^{\Qc }_{\Q \Qb}
+ \qqA^{\Qa }_{\Q \Qc} \qinv{\qqh}^{\Qc \Qd} \qqL^{\Q \Qe}_{\Qd }
\qqh^{}_{\Qe \Qb} - 2 \qqA^{\Qa }_{\Q \Qc} \qinv{\qqh}^{\Qc \Qd}
\qqalp^{\Q \Qe}_{\Qd } \qqh^{}_{\Qe \Qb} \dot \qT ,
 }
where
 \m { \qqalp^{\Q \Qe}_{\Qd } = \qinv{\qqJ}^{\QC }_{\QQ \Qd} \bit
 \qqalp^{\QQ \QD}_{\QC } {\qqJ}^{\Qe }_{\QQ \QD} },
according to the rules of transporting a material tensor to spacetime.

Within the range of elasticity, the relaxed metric could be flat, in
other words, it could make the material manifold a Euclidean affine
space. With thermal expansion, this no longer holds in general: The
relaxed metric makes the material manifold a curved Riemannian space.

Plasticity \1 2{see, e.g., \cite{RusRus11}} is another known source of
change of the relaxed structure, but, contrary to thermal expansion,
plastic changes are permanent. Related to the plastic deformation
originated change of the relaxed metric, the plastic change rate tensor
 \eq{@17789}{
\qqZ^{\Q \QB}_{\QA } := \frac {1}{2} \9 1 { \frac { \dd \qqg^{}_{\QA \QC}
}{\dd t} }_{\biTT\text{plastic}}^{\vphantom{|^|}} \qinv{\qqg}^{\QC
\QB}
 }
can be defined. Altogether then, we obtain
 \eq{@15448}{
{\dot{\qqA}}^{\Qa }_{\Q \Qb} = \qqL^{\Qa }_{\Q \Qc} \qqA^{\Qc }_{\Q \Qb}
+ \qqA^{\Qa }_{\Q \Qc} \qinv{\qqh}^{\Qc \Qd} \qqL^{\Q \Qe}_{\Qd }
\qqh^{}_{\Qe \Qb} - 2 \qqA^{\Qa }_{\Q \Qc} \qinv{\qqh}^{\Qc \Qd} \9 1 {
\qqalp^{\Q \Qe}_{\Qd } \dot \qT + \qqZ^{\Q \Qe}_{\Qd } } \qqh^{}_{\Qe
\Qb} .
 }

The small-deformedness regime is when the norm of the deformedness
tensor is small,
 \m { \9 7 { \9 3 { \qqD^{\Qa }_{\Q \Qb} } } \ll 1 }
and thus
 \m { \9 3 { \qqA^{\Qa }_{\Q \Qb} } = \exp \9 1 {2 \9 3 { \qqD^{\Qa
 }_{\Q \Qb} } } \approx \9 3 { \qqI^{\Qa }_{\Q \Qb} + 2 \qqD^{\Qa }_{\Q
 \Qb} } },
and then \re{@15448} leads, in the leading order of
 \m { \qqD^{\Qa }_{\Q \Qb} },
to
 \eq{@15991}{
\qsymm{\qqL}^{\Qa }_{\Q \Qb} = \dot\qqD^{\Qa }_{\Q \Qb} +
\qinv{\qqh}^{\Qa \Qc} \9 1 { \qqalp^{\Q \Qd}_{\Qc } \dot \qT + \qqZ^{\Q
\Qd}_{\Qc } } \qqh^{}_{\Qd \Qb} ,
 }
with \m {\bit \symm } standing for symmetric -- more closely, \m { \qqh
}-symmetric -- part, \m {
\qsymm{\qqL}^{\Qa }_{\Q \Qb} := \qfrac{1}{2} \2 2 { {\qqL}^{\Qa }_{\Q
\Qb} + \qinv\qqh^{\Qa \Qc} \qqL^{\Q \Qd}_{\Qc} \qqh^{}_{\Qd \Qb} } }.

Both sources of a changing relaxed metric, thermal expansion and plastic
processes, may ruin the flatness property of
 \m { \qqg^{}_{\QA \QB} }.
As an example, for elasticity plus thermal expansion one can find that,
in the small-deformedness regime, only temperature distributions with a
space independent gradient result in a zero Ricci tensor \1 2 {a classic
result \cite{HetEsl09}} while for large deformations, any nonhomegeneous
temperature distribution causes nonzero Ricci tensor \cite{Ful11} so the
compatibility condition for the relaxed metric is violated.

One can observe that we have used the relaxed metric as some kind of
reference quantity to define elastic shape and deformedness, but it
fundamentally differs from comparison to a reference configuration:
reference configuration is an auxiliary element -- involving choosing a
reference frame and a reference time -- which is not part of the
phenomenon to describe. In contradistinction, relaxed metric is a state
quantity, in other words, a physical field. The existence of this
additional field is what makes solids differ from fluids.

One could build thermomechanics with \m { \qqg^{}_{\QA \QB} } as one of
the state variables, but there are a few reasons to use \m { \qqA^{\QA
}_{\Q \QB} }, instead. First, the relaxed metric is not readily
accessible through measurements; second, already well-known elastic
energy expressions are more straightforward to make spacetime compatible
via \m { \qqA^{\QA }_{\Q \QB} }, and numerical calculations are also
expected to be more cumbersome in terms of the relaxed metric.

\section{Comparison to the usual kinematic framework}

Starting with the usual deformation gradient, its meaning can be freed
from reference frame but not from reference time. One can derive, with
time dependence emphasized,
 \eq{@18899}{
\qqF^{\Qa }_{\Q \Qb} \0 1 { t, t_0 } = \qqJ^{\Qa }_{\Q \QC} \0 1 { t }
\bittt \qinv{\qqJ}^{\QC }_{\QQ \Qb} \0 1 {t_0}
 \quad \text{and, for constant \m { \qqg^{}_{\QC \QD} },} \quad
\qqA^{\Qa }_{\Q \Qb} \0 1 { t } = \qqF^{\Qa }_{\Q \Qc} \0 1 { t, t_0 }
\bittt \qqA^{\Qc }_{\Q \Qd} \0 1 { t_0 } \qinv\qqh^{\Qd\Qe} \qqF^{\Qf}_{\Q
\Qe} \0 1 { t, t_0 } \qqh^{}_{\Qf\Qb}
 }
\1 2 {suppressing spatial (material point) dependence}. The first of
these equations shows how, by identifying material points \m { P } with
their spacetime position \m { \qr \1 1 { t_0, P }} at an instant \m
{t_0} chosen as reference time, the material manifold can be identified
with the Euclidean affine space of spacetime points at \m { t_0 }:
material vectors and tensors are mapped to spatial spacetime vectors and
tensors via \m { \qqJ^{\Qa }_{\bitttt \QC} \0 1 {t_0} }, and later world
line gradients \m { \qqJ^{\Qa }_{\Q \QC} \0 1 { t } } can be accessed as
\m { \qqJ^{\Qa }_{\Q \QC} \0 1 { t } = \qqF^{\Qa }_{\Q \Qb} \0 1 { t,
t_0 } \qqJ^{\Qb }_{\Q \QC} \0 1 {t_0} }. One danger is that thus one
also automatically -- and unnoticed -- transports the Euclidean \1 1
{i.e., flat Riemannian} structure of the Euclidean affine space of
spacetime points at \m { t_0 } to the material manifold. The result is
naturally the instantaneous metric \m { \qqh^{}_{\QC \QD} \1 1 { t_0 }
}, which is indeed a legitimate Riemann metric on the material manifold,
but is not the only one that may be relevant for constitutive and other
purposes. For example, comparison to \m { \qqh^{}_{\QC \QD} \1 1 { t_0 }
} implies that the continuum is considered undeformed at reference time,
and is supposed to fulfil the kinematic compatibility condition. This
assumption of an initial relaxed state at every material point is not
necessarily satisfied in practical applications. When opening an
underground tunnel, initial (in situ) stress is not zero; a laboratory
loading machine must apply some initial stress on the sample to keep it
firmly fixed; etc. Moreover, a generic plastic preceding history or any
inhomogeneous temperature distribution cause to violate the kinematic
compatibility condition mentioned above, the relaxed metric is not flat
so initial instantaneous metric -- a flat one by definition -- cannot be
equal to the relaxed one everywhere, generating elastic stress
[plasticity originated remanent (``frozen'') stress and unavoidable
thermal stress].

In parallel, various formulae emerging in continuum theory can require a
metric structure and, at each such situation, one should be able to
decide explicitly which metric, \m { \qqh^{}_{\QC \QD} } or \m {
\qqg^{}_{\QC \QD} }, is relevant at that situation. Such a decision
situation emerges for Fourier heat conduction, as we will find at
\re{@29315}.

Similarly to the assumption of an everywhere relaxed situation at \m
{t_0}, considering a homogeneous initial temperature distribution is
also artificial in general. This has consequences for thermal expansion
and thermal stresses.

The distinction between `right' and `left' vectors \1 1 {tensors, etc.}
is that `right' vectors are material vectors, tangent vectors of the
material manifold, while `left' vectors are spacelike spacetime vectors.
If, through \m { \qqJ^{\Qb }_{\Q \QC} \0 1 {t_0} }, one brings `right'
vectors/tensors to the same vector space in which the `left' ones live,
that may lead -- and indeed leads here and there -- to \m { t_0 }
dependent, hence, objectively forbidden, formulae.

It is also to be remembered that the classification `right' and `left'
leaves out non-spacelike spacetime vectors, which are also legitimate
objective quantities. In fact, similarly to how various physical areas
have been rewritten to be compatible with special and general
relativistic spacetime, continuum theory must also be possible to
rewrite to be totally compatible with Galilean, special and general
relativistic spacetime each. Such efforts, parallel to the one described
here, can be found, among others, at \cite{Van08,Van15,VanPavGrm15}.

The second formula of \re{@18899} gives, in the special case of zero
deformedness \1 1 {unit elastic shape tensor} at \m { t_0 },
 \m { \qqA^{\Qa }_{\Q \Qb} \0 1 { t } = \qqF^{\Qa }_{\Q \Qc} \0 1 { t,
 t_0 } \qinv\qqh^{\Qc\Qe} \qqF^{\Qf}_{\Q \Qe} \0 1 { t, t_0 }
 \qqh^{}_{\Qf\Qb} }
which tells that elastic shape generalizes the Cauchy--Green
tensor \1 1 {\m { \qqA^{\Qa }_{\Q \Qb} } the `left' one, and \m {
\qqA^{\QA }_{\Q \QB} } turns out to generalize the `right' one}.
Correspondingly, to obtain, for example, an objectively safe elastic
energy of the variable \m { \qqA^{\QA }_{\Q \QB} } \1 2 {or of \m {
\qqD^{\QA }_{\Q \QB} } like in \m { \frac {1}{2} \qqC^{\QA \Q \QC}_{\Q
\QB \QQ \QD} \qqD^{\QB }_{\Q \QA} \qD^{\QD }_{\QQ \QC} }} from a
usually used version written with the Cauchy--Green tensor, the
Cauchy--Green tensor variable is to be replace by the elastic shape
tensor.
 More generally, functions and formulae written in terms of \m {
\qqF^{\Qa }_{\Q \Qb} \0 1 { t, t_0 } } can be made spacetime compatible
if rewritable as functions of \m { t_0 } independent spacetime
compatible quantities seen above, via inserting at appropriate places
the silent assumption behind, \m { \qqI^{\Qa }_{\Q \Qb} = \qqA^{\Qa
}_{\Q \Qb} \0 1 { t_0 } }. Otherwise objectivity of their content is
questionable. During doing the rewriting, covectors should be
distinguished from vectors, and wherever a metric needs to be inserted,
a decision must be made whether to use the instantaneous metric or the
relaxed one.

Deformation and strain are change type quantities, somewhat like heat
and work in thermodynamics, in the sense that they express some
change occuring between \textit{two} instants, while elastic shape and
elastic deformedness are state describing quantities at \textit{one}
given instant. This conceptual difference explains why, for these new
quantities, new names are introduced.

Quantities like plastic deformation gradient may also fail to carry a
spacetime compatible meaning as they stand. It has to be a topic of
detailed future investigations how customary formulae of plasticity can
be brought into coherence with the approach of spacetime friendly
quantities.

\section{From mechanics to thermodynamics: isotropic case}\label{from}

Let us start from elastic mechanics. Until Sect.~\ref{ani}, let us
restrict ourselves to isotropic solids only. The customary balances for
mass and linear momentum are, fortunately, straightforward to rewrite as
frame free four-equations. Throughout the paper, only differential
equations in the bulk will be considered so it is just a side remark
here that, concerning boundary conditions, the switch to the spacetime
compatible quantities does not currently seem to require any
modification.
 The mechanical equations
 \eq{@21106}{
\dot \qrho = - \qrho \bit \nabla_{\Qa } \qqv^{\qf \Qa} ,
 \qquad
\qrho {\dot \qqv}^{\qf \Qa} = \nabla_{\Qb } \9 2 { \qinv{\qqh}^{\Qb \Qc}
\qqsig^{\Qa }_{\Q \Qc} } ,
 \qquad
{\dot{\qqA}}^{\Qa }_{\Q \Qb} = \qqL^{\Qa }_{\Q \Qc} \qqA^{\Qc }_{\Q \Qb}
+ \qqA^{\Qa }_{\Q \Qc} \qinv{\qqh}^{\Qc \Qd} \qqL^{\Q \Qe}_{\Qd }
\qqh^{}_{\Qe \Qb}
 }
together with a constitutively known
 \m { \qqsig^{\Qa }_{\Q \Qb} = \qqsig^{\Qa }_{\Q \Qb} \0 1 { \0 3
 {\qqA^{\Qc }_{\Q \Qd}} } },
or, equivalently, 
 \m { \qqsig^{\Qa }_{\Q \Qb} = \qqsig^{\Qa }_{\Q \Qb} \0 1 { \0 3
 {\qqD^{\Qc }_{\Q \Qd}} } },
form a closed set of equations. As an initial value problem, the initial
distribution of \m { \qqA^{\Qa }_{\Q \Qb} } is arbitrary as long
as \m { \qqg^{}_{\QC \QD} = \qqh^{}_{\QC \QE} \qqA^{\QE}_{\Q\Q \QD} } is
flat, and later evolution of \m { \qqA^{\Qa }_{\Q \Qb} } happens
according to the velocity field. In other words, the time evolution
equation for \m { \qqA^{\Qa }_{\Q \Qb} } is equivalent to that \m {
\qqg^{}_{\QC \QD} } is time independent. The time evolution equation form
will become practical at later stages.
 For the purpose of
 \m { \qqsig^{\Qa }_{\Q \Qb} = \qqsig^{\Qa }_{\Q \Qb} \0 1 { \0 3
 {\qqD^{\Qc }_{\Q \Qd}} } },
we may think, for example, of
 \eq{@21340}{
\qqsig^{\Qa }_{\Q \Qb} = \qE\dev \0 1 {\qqD\dev}\QI^{\Qa }_{\Q \Qb} +
\qE\sph \0 1 {\qqD\sph}\QI^{\Qa }_{\Q \Qb}
 \qquad
\9 1 { \qE\dev = 2G, \quad  \qE\sph = 3K } .
 }
Note that \m { \qqsig^{\Qa }_{\Q \Qb} } is not directly symmetric but is
to be \m { \qqh_{\Qc \Qd} }--symmetric, like seen for \m {
\qqA^{\Qa }_{\Q \Qb} } and \m { \qqD^{\Qa }_{\Q \Qb} }.

Elasticity assumes, further, the existence of a specific elastic energy
 \m { \qel\qe \2 1 { \0 3 { \qqD^{\Qa }_{\Q \Qb} } } },
an isotropic scalar function of
 \m { \0 3 { \qqD^{\Qa }_{\Q \Qb} } },
with the properties
 \eq{@21818}{
\qqsig^{\Qa }_{\Q \Qb} = \qrho \frac {\dd \qel\qe}{\dd \qqD^{\Qb
}_{\Q \Qa}} ,
 \qquad
\qrho \qel{\dot{\qe}} = \qqsig^{\Qa }_{\Q \Qb} \qsymm{\qqL}^{\Qb
}_{\Q \Qa} .
 }
Here, the second formula can be proved from the first one and from
\re{@13637} \1 1 {its `left' version} as follows:
 \eq{@2219}{
\qrho \qel{\dot{\qe}} = \qrho {\frac {\dd \qel\qe}{\dd \qqA^{\Qa }_{\Q
\Qb} }} {\dot{\qqA}}^{\Qa }_{\Q \Qb} & = \qrho {\frac {\dd \qel\qe}{\dd
\qqA^{\Qa }_{\Q \Qb} }} \9 2 { \qqL^{\Qa }_{\Q \Qc} \qqA^{\Qc }_{\Q \Qb}
+ \qqA^{\Qa }_{\Q \Qd} \qinv{\qqh}^{\Qd \Qe} \qqL^{\QQ \Qf}_{\Qe }
\qqh^{}_{\Qf \Qb} }
 \tag*{} \\ \label{@22110}
& = \qrho \qqA^{\Qc }_{\Q \Qb} \frac {\dd \qel\qe}{\dd \qqA^{\Qa }_{\Q
\Qb} } \qqL^{\Qa }_{\Q \Qc} + \qrho \frac {\dd \qel\qe}{\dd \qqA^{\Qa
}_{\Q \Qb} } \qqA^{\Qa }_{\Q \Qd} \9 2 { \qinv{\qqh}^{\Qd \Qe} \qqL^{\QQ
\Qf}_{\Qe } \qqh^{}_{\Qf \Qb} } = \qqsig^{\Qc }_{\Q \Qa}
\qsymm{\qqL}^{\Qa }_{\Q \Qc} .
 }
Apparently, this simple result relies on the logarithmic definition of
elastic deformedness.

It is interesting to observe that, by \re{@21818}, we have also obtained
 \m{ \qqsig^{\Qa }_{\Q \Qb} {\dot{\qqD}}^{\Qb }_{\Q \Qa} = 
 \qqsig^{\Qa }_{\Q \Qb} \qsymm{\qqL}^{\Qb}_{\Q \Qa} , }
since
 \eq{@23680}{
\qqsig^{\Qa }_{\Q \Qb} \qsymm{\qqL}^{\Qb}_{\Q \Qa} = \qrho
\qel{\dot{\qe}} = \qrho \frac {\dd \qel\qe}{\dd \qqD^{\Qb }_{\Q \Qa}}
{\dot{\qqD}}^{\Qb }_{\Q \Qa} = \qqsig^{\Qa }_{\Q \Qb} {\dot{\qqD}}^{\Qb
}_{\Q \Qa} .
 }
However,
 \m{ \qqsig^{\Qa }_{\Q \Qb} {\dot{\qqD}}^{\Qb }_{\Q \Qa} =
 \qqsig^{\Qa }_{\Q \Qb} \qsymm{\qqL}^{\Qb}_{\Q \Qa} }
holds not because
 \m { \qsymm{\qqL}^{\Qb}_{\Q \Qa} }
itself equals
 \m { {\dot{\qqD}}^{\Qb }_{\Q \Qa} }
\1 1 {that would be true only in the small-deformedness regime} but
because of some much less trivial reasons \1 1 {including isotropy}.

Thermal effects enforce elasticity to be generalized from various
aspects. On one side,
 \m{ \qel\qe = \qel\qe \0 1 { \qT, \9 3 { \qqD^{\Qa}_{\Q \Qb} } } }
in general \1 2 {e.g.,
 in \re{@21340},
\m { \qE\dev = \qE\dev \1 1 { \qT } }, \m { \qE\sph = \qE\sph \1 1 { \qT
} }}. Second, there may be thermal expansion:
 \m{ \qqg^{}_{\QC \QD} = \qqg^{}_{\QC \QD} (\qT)}.
Third, in addition to mechanical power, a heat flux
 \m { \0 1 { \qqj\biT_\qe }^{\Qa} }
may also be present. Then, having the first law of thermodynamics in
mind, we expect
 \m { \qrho \qel{\dot{\qe}} = \qqsig^{\Qa }_{\Q \Qb} \qsymm{\qqL}^{\Qb  
 }_{\Q \Qa} }
to be generalized to a balance
 \eq{@23831}{
\qrho \dot{\qe} = - \nabla^{}_{\Qa} \0 1 { \qqj\biT_\qe }^{\Qa} +
\qqsig^{\Qa }_{\Q \Qb} \qsymm{\qqL}^{\Qb }_{\Q \Qa} .
 }
But with what
 \m { \qe \0 1 { \qT, \9 3 { \qqD^{\Qa}_{\Q \Qb} } } } and \m {
 \0 1 { \qqj\biT_\qe }^{\Qa}
 }? In parallel, on
thermodynamical grounds, we anticipate the existence of a specific
entropy
 \m { \qs \0 1 { \qT, \9 3 { \qqD^{\Qa}_{\Q \Qb} } } }
as well, with a balance
 \eq{@24144}{
\qrho \dot{\qs} = - \nabla^{}_{\Qa} \0 1 { \qqj\biT_\qs }^{\Qa} + \qpi_\qs
 }
where the source term \m { \qpi_\qs } -- entropy production -- is
expected to be only heat-related since elasticity and thermal expansion
are experienced as reversible phenomena. Taking a look at nonequilibrium
thermodynamics, we can assume
 \m{ \0 1 { \qqj\biT_\qs }^{\Qa} = \frac {1}{\qT} \0 1 { \qqj\biT_\qe
 }^{\Qa} },
but what
 \m { \qs \0 1 { \qT, \9 3 { \qqD^{\Qa}_{\Q \Qb} } } }
would fulfil our hope?

Let us start with a calculation generalizing
\re{@22110}, and utilizing \re{@18040}:
 \eq{@2110}{
\qrho \qel{\dot{\qe}} = \qrho \frac {\pd \qel\qe}{\pd \qT } \dot\qT +
\qrho \frac {\pd \qel\qe}{\pd \qqA^{\Qa }_{\Q \Qb} } {\dot{\qqA}}^{\Qa
}_{\Q \Qb} = \qrho \frac {\pd \qel\qe}{\pd \qT } \dot\qT + \qqsig^{\Qc
}_{\Q \Qa} \qsymm{\qqL}^{\Qa}_{\Q \Qc} - \qqsig^{\Qb }_{\Q \Qc}
\qinv{\qqh}^{\Qc \Qd} \qqalp^{\Q \Qe}_{\Qd } \qqh^{}_{\Qe \Qb} \dot \qT
.
 }
Now let us subtract this from the balance \re{@23831}, and aim at
forming \re{@24144} multiplied by \m { \qT }:
 \eq{@210}{
\qrho \9 1 { \qe - \qel\qe }\qdot = - \nabla^{}_{\Qa} \0 1 { \qT
\qqj\biT_\qs }^{\Qa} - \qrho \frac {\pd \qel\qe}{\pd \qT } \dot\qT & +
\qqsig^{\Qb }_{\Q \Qc} \qinv{\qqh}^{\Qc \Qd} \qqalp^{\Q \Qe}_{\Qd }
\qqh^{}_{\Qe \Qb} \dot \qT ,
 \\ \label{jbds}
\qrho \9 1 { \qT \frac {\qe - \qel\qe}{\qT} }\qdot + \9 1 { \qrho \frac
{\pd \qel\qe}{\pd \qT } - \qqsig^{\Qb }_{\Q \Qc} \qinv{\qqh}^{\Qc \Qd}
\qqalp^{\Q \Qe}_{\Qd } \qqh^{}_{\Qe \Qb} } \dot\qT & = - \qT \bitt
\nabla^{}_{\Qa} \0 1 { \qqj\biT_\qs }^{\Qa} - \9 1 { \nabla^{}_{\Qa} \qT
} \0 1 { \qqj\biT_\qs }^{\Qa} ,
 \\ \label{jbdsr}
\qrho \qT\9 1 { \frac {\qe - \qel\qe}{\qT} }\qdot + \qrho \4 2 { \qe -
\qel\qe + \qT \4 1 { \frac {\pd \qel\qe}{\pd \qT } - \frac {1}{\qrho}
\qqsig^{\Qb }_{\Q \Qc} \qinv{\qqh}^{\Qc \Qd} \qqalp^{\Q \Qe}_{\Qd }
\qqh^{}_{\Qe \Qb} } } \frac{\dot\qT}{\qT} & =
- \qT \bitt \nabla^{}_{\Qa} \0 1 { \qqj\biT_\qs }^{\Qa}
- \9 1 { \nabla^{}_{\Qa} \qT } \frac {1}{\qT} \0 1 { \qqj\biT_\qe }^{\Qa} ,
 \\ \label{jbd}
\qrho \qT\9 1 { \frac {\qe - \qel\qe}{\qT} }\qdot + \qrho \qth\qe \frac
{\dot\qT}{\qT} & = - \qT \bitt \nabla^{}_{\Qa} \0 1 { \qqj\biT_\qs
}^{\Qa} + \qT \0 1 { \nabla^{}_{\Qa} \frac{1}{\qT} } \0 1 { \qqj\biT_\qe
}^{\Qa}
 }
with
 \eq{@27130}{
\qth\qe = \qth\qe \0 1 { \qT, \9 3 { \qqD^{\Qa}_{\Q \Qb} } } = \qe -
\qel\qe + \qT \0 1 { \frac {\pd \qel\qe}{\pd \qT } - \frac {1}{\qrho}
\qqsig^{\Qb }_{\Q \Qc} \qinv{\qqh}^{\Qc \Qd} \qqalp^{\Q \Qe}_{\Qd }
\qqh^{}_{\Qe \Qb} } .
 }
This is of the form \re{@24144} multiplied by \m {
\qT }, where the would-be entropy production is related to heat only (no
elastic or thermal expansion contribution). The only snag is the second
term on the lhs of \re{jbd}: it also should be of the form
 \m{ \qrho \qT \1 1 {\text{\textit{something}}}\qdot . }
This can be satisfied if \m { \qth\qe = \qth\qe \1 1 { \qT } }.
Then, writing \re{jbd} as
 \eq{@27717}{
\qrho \qT \9 2 { \frac {\qth\qe}{\qT} - \0 1 { \frac {\pd \qel\qe}{\pd \qT } -
\frac {1}{\qrho} \qqsig^{\Qb }_{\Q \Qc} \qinv{\qqh}^{\Qc \Qd} \qqalp^{\Q
\Qe}_{\Qd } \qqh^{}_{\Qe \Qb} } }\qdot + \qrho \qth\qe \frac {\dot\qT}{\qT}
& = - \qT \bitt \nabla^{}_{\Qa} \0 1 { \qqj\biT_\qs }^{\Qa} + \qT \0 1 {
\nabla^{}_{\Qa} \frac{1}{\qT} } \0 1 { \qqj\biT_\qe }^{\Qa} ,
 }
and rearranging this as
 \eq{poqw}{
\qrho \qT \9 1 { \frac {1}{\qrho} \qqsig^{\Qb }_{\Q \Qc}
\qinv{\qqh}^{\Qc \Qd} \qqalp^{\Q \Qe}_{\Qd } \qqh^{}_{\Qe \Qb} - \frac
{\pd \qel\qe}{\pd \qT } }\qdot + \qrho \qT \9 2 { \9 1 { \frac
{\qth\qe}{\qT} }\qdot + \frac {\qth\qe}{\qT^2} \dot\qT } =  - \qT \bitt
\nabla^{}_{\Qa} \0 1 { \qqj\biT_\qs }^{\Qa} + \qT \0 1 { \nabla^{}_{\Qa}
\frac{1}{\qT} } \0 1 { \qqj\biT_\qe }^{\Qa} ,
 }
we have
 \eq{@28446}{
\9 1 { \frac {\qth\qe}{\qT} }\qdot + \frac {\qth\qe}{\qT^2} \dot\qT & =
\frac {\dd \qth\qe}{\dd \qT } \frac{\dot\qT}{\qT} = {\qth\qs \1 1 { \qT
} }\qdot = \frac {\dd \qth\qs}{\dd \qT } \dot\qT
 \qquad \text{with} \qquad
\frac {\dd \qth\qs}{\dd \qT } = \frac{1}{\qT} \frac {\dd \qth\qe}{\dd \qT } .
 }
Then we can read off
 \eq{@25531}{
\qe & = \qth\qe \1 1 {\qT} + \qel\qe + \qT \9 1 { \frac {1}{\qrho}
\qqsig^{\Qb }_{\Q \Qc} \qinv{\qqh}^{\Qc   \Qd} \qqalp^{\Q \Qe}_{\Qd }
\qqh^{}_{\Qe \Qb} - \frac {\pd \qel\qe}{\pd \qT } } ,
 \\ \label{dcjsdc}
\qs & = \qth\qs \1 1 {\qT} + \9 1 { \frac {1}{\qrho} \qqsig^{\Qb }_{\Q
\Qc} \qinv{\qqh}^{\Qc \Qd} \qqalp^{\Q \Qe}_{\Qd } \qqh^{}_{\Qe \Qb} -
\frac {\pd \qel\qe}{\pd \qT } } ,
 \qquad
 \qquad
\qpi_\qs = \0 1 { \nabla^{}_{\Qa} \frac{1}{\qT} } \0 1 {
\qqj\biT_\qe }^{\Qa} .
 }
The term \m { \qth\qe \1 1 {\qT} } is related to
specific heat, \m { c|_{\qqD^{\Qa}_{\bitttt \Qb} = 0} }, while the last
equation here
is the entropy production which we can ensure to be
positive definite via, e.g., Fourier heat conduction, for which both
forms
 \eq{@29315}{
\0 1 { \qqj\biT_\qe }^{\Qa} = \qlam \qinv{\qqg}^{\Qa \Qb}
\0 0 { \nabla^{}_{\Qb} \frac{1}{\qT} }
 \qquad  \mbox{or}  \qquad
\0 1 { \qqj\biT_\qe }^{\Qa} = \qlam \qinv{\qqh}^{\Qa \Qb}
\0 0 { \nabla^{}_{\Qb} \frac{1}{\qT} }
 }
are allowed \1 1 {with a non-negative heat conduction coefficient \m {
\qlam }}. The former, i.e., choosing the relaxed metric for connecting
the covector gradient with the vector lhs, is more plausible physically
but it must be a topic of further study what metric to choose here.

Note that, in \re{@25531}--\re{dcjsdc}, stress is purely elastic and may
well be
 \m{ \qqsig^{\Qa }_{\Q \Qb} = \qqsig^{\Qa }_{\Q \Qb} 
 \0 1 { \9 3 { \qqD^{\Qc}_{\Q \Qd} } } }
(no \m { \qT } dependence). Nevertheless, the terms coupling \m { \qT }
and
 \m { \qqD^{\Qc}_{\Q \Qd} }
are sources of thermal stress and of Joule--Thomson effect. As a special
limiting case, small-strain Duhamel--Neumann thermoelasticity can
be obtained \cite{FulVanCsa13Bre}.

When incorporating plastic changes, the only modification is an
additional entropy production term
 \m { \frac {1}{\qT} \qqsig^{\Qb }_{\Q \Qc} \qinv{\qqh}^{\Qc \Qd}
 \qqZ^{\Q \Qe}_{\Qd } \qqh^{}_{\Qe \Qb} }.
Its positive definiteness can be ensured, for example, with the simple
yet plausible
 \eq{@42180}{
\qqZ^{\bitttt \Qb}_{\Qa } = \Gamma \bit \qqh^{}_{\Qa\Qc}
\0 1 {{\dot\qqsig}\dev}^{\Qc}_{\QQ \Qd} \qinv{\qqh}^{\Qd \Qb}
 \quad \text{with} \quad
\Gamma = \gamma H \0 1 { \0 1 {\qqsig\dev}^{\Qa}_{\Q \Qb} \0 1
{\qqsig\dev}^{\Qb}_{\Q \Qa} - \qfrac {2}{3} \qsig^{2}_{\text{yield}}}
\bitt H \0 1 { \0 1 {\qqsig\dev}^{\Qa}_{\Q \Qb} \0 1
{{\dot\qqsig}\dev}^{\Qb}_{\Q \Qa} } ,
 \quad \1 0 { \gamma > 0 } ,
 }
the first Heaviside function \m { H } embodying von Mises type yield
criterion. Note how irreversible thermodynamics switches off plastic
change during unloading \1 1 {second Heaviside function}: Entropy
production is not allowed to be negative.

\section{Anisotropy}\label{ani}

Anisotropy means distinguished \textit{material} directions.
In a linear elastic model, for example, it is not
 \m { \qqsig^{\Qa }_{\Q \Qb} } and \m { \qqD^{\Qc }_{\Q \Qd} }
but their material version
 \m { \qqsig^{\QA }_{\Q \QB} } and \m { \qqD^{\QC }_{\QQ \QD} }
between which the linear coefficient tensor \m { \qqC^{\QA \QQ \QD}_{\Q
\QB \QQ \QC} } is constant, in
 \m { \qqsig^{\QA }_{\Q \QB} = \qqC^{\QA \QQ \QD}_{\Q \QB \QQ \QC}
 \qqD^{\QC }_{\QQ \QD} }.
 Other constitutive properties are also expected to be
connected to the material
form rather than to the spacelike spacetime form. \1 1 {In the meantime,
balances primarily live on spacetime.}

Let us start with the material version of \re{@15448}, brought to a form
that comes useful later:
 \eq{@45158}{
{\dot{\qqA}}^{\QA }_{\Q \QB} & = - \qinv{\qqg}^{\QA \QC}
{\dot\qqg}^{}_{\QC \QD} \qinv{\qqg}^{\QD \QE} \qqh^{}_{\QE \QB} + 2
\qqA^{\QA }_{\Q \QC} \qsymm{\qqL}^{\QC}_{\QQ \QB}
 \tag*{} \\ \label{iocds}
& = - 2 \qqA^{\QA }_{\Q \QE} \qinv{\qqJ}^{\QE}_{\Q\Q \Qa}
\qinv{\qqh}^{\Qa \Qb} \9 1 { \qqalp^{\Q \Qc}_{\Qb } \dot \qT + \qqZ^{\Q
\Qc}_{\Qb } } \qqh^{}_{\Qc \Qd} \qqJ^{\Qd}_{\Q \QB} + 2 \qqA^{\QA }_{\Q
\QC} \qsymm{\qqL}^{\QC}_{\QQ \QB} .
 }
Next, we must realize that no part of \re{@13637} may hold for no
isotropy, including that the factors in the products are not commuting
with one another. Though we still expect an elastic energy to exist: Is
it related to elastic stress according to the first, the second or the
third version in \re{@13637}? It is the energy balance in which they
should have a distinguished relationship so let us investigate the
anisotropic version of \re{@210}:
 \eq{@46356}{
\qrho \0 1 { \qe - \qel\qe }\qdot = - \nabla^{}_{\Qa} \0 1 {
\qqj\biT_\qe }^{\Qa} & + \qqsig^{\Qa }_{\bittt \Qc} \qsymm{\qqL}^{\Qc }_{\Q
\Qa} - \qrho \frac {\pd \qel\qe}{\pd \qT } \dot\qT - \qrho \frac {\pd
\qel\qe}{\pd \qqA^{\QA }_{\bitttt \QB} } {\dot{\qqA}}^{\QA }_{\bitttt \QB}
 = - \nabla^{}_{\Qa} \0 1 { \qqj\biT_\qe }^{\Qa} + \qqsig^{\QB }_{\Q\QC}
\qsymm{\qqL}^{\QC }_{\Q \QB} - 2 \qrho \frac {\pd \qel\qe}{\pd \qqA^{\QA
}_{\bitttt \QB} } \qqA^{\QA }_{\bitttt \QC}
\qsymm{\qqL}^{\QC}_{\Q \QB }
 \tag*{} \\ \label{dbwkh}
& \quad + 2 \qrho \frac {\pd \qel\qe}{\pd \qqA^{\QA }_{\Q \QB} } \qqA^{\QA
}_{\Q \QE} \qinv{\qqJ}^{\QE}_{\Q\Q \Qa} \qinv{\qqh}^{\Qa \Qb} \9 1 {
\qqalp^{\Q \Qc}_{\Qb } \dot \qT + \qqZ^{\Q \Qc}_{\Qb } } \qqh^{}_{\Qc
\Qd} \qqJ^{\Qd}_{\Q \QB} - \qrho \frac {\pd \qel\qe}{\pd \qT } \dot\qT
 \tag*{} \\ \label{edbwkedew}
& = - \nabla^{}_{\Qa} \0 1 { \qqj\biT_\qe }^{\Qa} + \9 1 { \qqsig^{\QB
}_{\Q\QC} - 2 \qrho \frac {\pd \qel\qe}{\pd \qqA^{\QA }_{\Q \QB} }
\qqA^{\QA }_{\Q \QC} } \qsymm{\qqL}^{\QC}_{\Q\Q \QB }
 \tag*{} \\ \label{edbwk}
& \quad + \9 2 { \qqJ^{\Qd}_{\Q \QB} \9 1 { 2 \qrho \frac {\pd \qel\qe}{\pd
\qqA^{\QA }_{\Q \QB} } \qqA^{\QA }_{\Q \QC} } \qinv{\qqJ}^{\QC}_{\Q\Q
\Qa} } \qinv{\qqh}^{\Qa \Qb} \9 1 { \qqalp^{\Q \Qc}_{\Qb } \dot \qT +
\qqZ^{\Q \Qc}_{\Qb } } \qqh^{}_{\Qc \Qd} - \qrho \frac {\pd \qel\qe}{\pd
\qT } \dot\qT .
 }
Then we can see that, with the choice
 \eq{@47640}{
\qqsig^{\QB }_{\Q\QC} = 2 \qrho \frac {\pd \qel\qe}{\pd \qqA^{\QA }_{\Q
\QB} } \qqA^{\QA }_{\Q \QC} ,
 }
two terms become simple, and \re{edbwk} reduces to
 \eq{@4356}{
\qrho \9 1 { \qe - \qel\qe }\qdot = - \nabla^{}_{\Qa} \0 1 {
\qqj\biT_\qe }^{\Qa} + \qqsig^{\Qd }_{\Q \Qa} \qinv{\qqh}^{\Qa \Qb} \9 1
{ \qqalp^{\Q \Qc}_{\Qb } \dot \qT + \qqZ^{\Q \Qc}_{\Qb } } \qqh^{}_{\Qc
\Qd} - \qrho \frac {\pd \qel\qe}{\pd \qT } \dot\qT .
 }
This is analogous to \re{@210} so, from here, we can proceed the same
way, and the result will also be \re{@25531}--\re{dcjsdc} \1 2 {the
entropy production being extended by the same plasticity related term
as in the isotropic case}.

Therefore, after settling \re{@47640}, nothing differs from the
isotropic formulae. Note that, if we had used the opposite product order
in \re{@17071} and \re{@17789}, the calculation would not have led to
such a nice outcome.

As a summary, the objective thermomechanics obtained here works with
basic fields \m { \qr^{\qf \Qc} }, \m { \qqA^{\QC }_{\QQ \QD} }, \m {
\qrho } and \m { \qT } as functions of \m { \1 1 { t, P } }, with
derived fields \m { \qqv^{\qf \Qc} }, \m { \qqJ^{\Qc}_{\Q \QC} }, \m {
\qqL^{\Qc}_{\Q \Qd} } and \m { \qqh^{}_{\QC \QD} }, with basic
constitutive functions \m { \qth\qe \1 1 {\qT} }, \m { \qel\qe \0 1 {
\qT, \9 3 { \qqD^{\Qa}_{\bitttt \Qb} } } }, \m { \qqalp^{\Q\Q \QD}_{\QC}
\1 1 { \qT } }, \m { \qqZ^{\Q\Q \QD}_{\QC } } and \m { \0 1 {
\qqj\biT_\qe }^{\QC} } \1 2 {variables for the latter two can be written
in various ways} and derived constitutive functions \m { \qqsig^{\QC
}_{\QQ \QD} \0 1 { \qT, \9 3 { \qqD^{\Qa}_{\bitttt \Qb} } } }, \m { \qs
\0 1 { \qT, \9 3 { \qqD^{\Qa}_{\bitttt \Qb} } } } and \m { \0 1 {
\qqj\biT_\qs }^{\QC} }.

Apparently, this formulation is not a thermodynamically elegant one, but
is at least a thermodynamically consistent as well as objective one, and
the special cases of a constant specific heat, thermal expansion
coefficient, heat conduction coefficient and elasticity coefficients \m
{2G }, \m { 3K } -- important for many engineering applications -- can
be explicitly expressed. Naturally, more elegant reformulations are
worth exploring.

\section{Discussion and outlook}\label{concl}

The framework presented here guarantees objectivity via the frame free
spacetime formulation, and describes elasticity, thermal expansion and
plasticity in a thermodynamically consistent theory. Naturally, there
are various tasks for the future. Continuing comparison with the
literature, especially on the plasticity side, is necessary, both for
the kinematic quantities and on the constitutive content.
Rheology/viscoelasticity is to be incorporated -- here, the tensorial
internal variable methodology realized so far for isotropic solids and
small deformations \cite{AssFulVan15} seems to pose no problem against
generalization, and the possibility of nonequilibrium thermodynamical
coupling of this tensorial phenomenon to plasticity promises interesting
predictions. This may still not be enough to describe all
large-deformation complex rheological phenomena like considered in
\cite{PalVar10}, but the number of internal variables can be increased
and other possibilities are also to be investigated.

Further phenomena like damage and failure \cite{Van96,VanVas01} are to
be added, too. To explore the spacetime aspects of GENERIC and
other nonequilibrium thermodynamical frameworks is also an important
mission.

The frame free approach can be advantageous not only because it helps
avoiding artifacts and mistakes: It also catalyses and even enforces
better physical understanding. For example, the notion of the relaxed
metric was born motivated by the urgent need of distinguishing solids
from fluids. When discussing other thermodinamical four-quantities like
four-energy-momentum \cite{VanPavGrm15} and related issues, further
similar outcomes and discoveries can be expected.

\begin{acknowledgement}
The author thanks Csaba Asszonyi, Tam\'as Matolcsi and P\'eter V\'an for
many valuable discussions.
\end{acknowledgement}

\def\acknowledgementname{Funding}
\begin{acknowledgement}
This work was supported by the grants OTKA K82024 and K116375.
\end{acknowledgement}


\end{document}